# KM-UNet: KAN Mamba UNet for medical image segmentation

Yibo Zhang[a], Jingwen Zhao[a*], Xiang Liu[a], Xian Tang[a], Yunyu Shi[a], Lina Wei[b], Guyue Zhang[c]
[a]Shanghai University of Engineering Science
[b]Hangzhou City University
[c]Zhejiang Academy of Quality Science

## Abstract

Medical image segmentation is a critical task in medical imaging analysis. Traditional CNN-based methods struggle with modeling long-range dependencies, while Transformer-based models, despite their success, suffer from quadratic computational complexity. To address these limitations, we propose KM-UNet, a novel U-shaped network architecture that combines the strengths of Kolmogorov-Arnold Networks (KANs) and state-space models (SSMs). KM-UNet leverages the Kolmogorov-Arnold representation theorem for efficient feature representation and SSMs for scalable long-range modeling, achieving a balance between accuracy and computational efficiency.

We evaluate KM-UNet on five benchmark datasets: ISIC17, ISIC18, CVC, BUSI, and GLAS. Experimental results demonstrate that KM-UNet achieves competitive performance compared to state-of-the-art methods in medical image segmentation tasks.

To the best of our knowledge, KM-UNet is the first medical image segmentation framework integrating KANs and SSMs. This work provides a valuable baseline and new insights for the development of more efficient and interpretable medical image segmentation systems. The code is open source at https://github.com/2760613195/KM_UNet

## 1. Introduction

In the past decade, significant progress has been made in medical image segmentation methods to meet the needs of computer-aided diagnosis and image-guided surgery systems [1][2][3]. U-Net[4] is a landmark in this field, demonstrating for the first time the effectiveness of encoder-decoder convolutional networks with skip connections in medical image segmentation[5][6]. Subsequently, a series of improvements to U-Net have emerged, such as U-Net++[7], 3D U-Net, and V-Net[8], which further expanded its application scope and performance. Moreover, hybrid architectures like U-NeXt combine convolutional operations with MLP to enhance the efficiency of segmentation networks, especially in resource-constrained environments.

At the same time, Transformer-based networks have gained attention for their ability to model long-range dependencies and global context[9][10]. For example, TransUNet[11] and Swin-UNet[12] leverage Vision Transformers (ViT)[13] and Swin Transformers[14], demonstrating powerful global modeling capabilities. However, despite their strong performance, these models demand high data volumes and computational resources, posing challenges in data-limited or real-time processing scenarios[15].

In recent years, structured state-space models (SSMs) have provided an efficient solution for modeling long sequences with linear computational complexity[16][18]. Models such as U-Mamba[17] and SegMamba[19] have shown performance improvements by combining SSMs with CNNs in medical image segmentation. However, despite these innovations, existing U-Net variants still face fundamental challenges in kernel design and black-box attributes, which affect model interpretability and diagnostic reliability.

To address these challenges, this paper proposes a novel U-Net architecture, KM-UNet, which integrates Kolmogorov-Arnold Networks (KAN) and Selective-Scan Efficient Multi-scale (SEM) attention modules. KM-UNet consists of an encoder, a decoder, and skip connections. The encoder utilizes the SEM attention module for feature extraction, combined with patch merging operations for downsampling. The decoder incorporates the SEM attention module and patch expansion operations to restore the segmentation results to their original size. In the skip connections, we use simple addition to highlight the segmentation performance of the pure SSM model.With theoretical support from the Kolmogorov-Arnold equation, the KAN network provides inherent interpretability, significantly enhancing the network's transparency. This advantage bridges the gap between the network's physical properties and its actual performance, making the decision-making process of the model easier to understand and analyze. On the other hand, the Mamba module, inspired by Structured State-

Space Modeling (SSM), enhances the network's ability to capture complex nonlinear spatial dependencies and long-range interactions.

By combining the strengths of both techniques, we are able to overcome the limitations of traditional convolutional neural networks (CNN) and Transformer architectures in handling complex nonlinear spatial dependencies and long-range interactions. The high interpretability of KAN ensures model transparency and clinical trustworthiness, while the Mamba module resolves the trade-off between long-range dependency capture and computational efficiency. This innovative integration enables KM-UNet to achieve outstanding performance in medical image segmentation tasks, improving not only segmentation accuracy but also the reliability and trustworthiness of the model in practical applications.

The main contributions of this paper are summarized as follows:

1) We presents the KM-UNet model, exploring the potential of the KAN-SSM fusion model in the field of medical image segmentation, offering valuable insights for the development of more efficient KAN-based segmentation methods.

2) The paper also introduces the Selective Scanning Efficient Multi-Scale (SEM) attention module, which enables multi-scale information learning across space and cleverly integrates the concept of the SSM model.

3) We propose a novel scanning strategy that performs a rotation from the outer layers toward the center, which allows for capturing features that are difficult to detect with traditional scanning methods. This approach enhances accuracy, particularly for single-object tasks or segmentation tasks involving simpler shapes.

## 2. Related Work

### Research on U-Net Architecture Based on KAN Network for Medical Image Segmentation

Medical image segmentation has long been a research hotspot in both computer vision and healthcare due to its crucial role in clinical diagnosis and treatment[21][22]. U-Net, as a classic encoder-decoder architecture, has achieved significant breakthroughs in medical image segmentation by effectively capturing multi-scale features through skip connections. However, traditional CNN models, limited by their local receptive fields, struggle to capture global information adequately, leading to insufficient feature extraction and limited segmentation performance when processing complex images.

The introduction of Kolmogorov–Arnold Network (KAN) provides a new perspective to address these issues. The KAN network, with its highly interpretable structural design, compensates for the shortcomings of CNN models in modeling complex nonlinear relationships [20]. Its modular construction not only enhances the model's global context understanding but also establishes a strong connection between physical characteristics and empirical performance. In the field of medical image segmentation, combining the KAN network with U-Net not only leverages the multi-scale feature fusion advantage of U-Net but also improves the network's interpretability and global information capture capability, which is expected to enhance the model's stability and reliability in complex medical image segmentation tasks.[35]

### Integration of Mamba Module with U-Net for Enhanced Medical Image Segmentation

In recent years, the application of State Space Models (SSM) in long sequence modeling has gained significant attention, with modern SSMs, such as Mamba [23], demonstrating excellent performance in visual tasks due to their linear time complexity and efficient training process. Models like U-Mamba [17] have successfully combined SSM with CNNs, applying them for the first time in medical image segmentation tasks, showcasing their potential for handling complex anatomical structures. SegMamba [19] introduced the SSM module in the encoder while keeping CNNs in the decoder, creating a hybrid model for 3D medical image segmentation tasks, such as brain tumor segmentation, and achieving promising results. Integrating the Mamba module into the U-Net architecture enables the network to capture long-range dependencies and multi-scale features with minimal computational overhead, enhancing U-Net's global modeling capability and segmentation accuracy[36]. This integration lays a solid foundation for further advancements in efficient medical image segmentation methods. Additionally, multi-stream networks like PSA (Parallel Spatial Attention) improve feature representation in visual tasks by modeling long-range dependencies in parallel, though they introduce extra computational cost. Furthermore, while the triple attention mechanism enhances feature abstraction by combining cross-channel and spatial information, its simple averaging strategy for weight aggregation limits the discriminative power of deep features. On the other hand, multi-scale convolutions, such as those in Inception networks and selective convolutional networks, improve feature representation by adapting the receptive field through multi-branch structures. Our proposed multi-scale attention module optimizes global context capturing by leveraging cross-space learning during feature fusion, further improving segmentation performance.

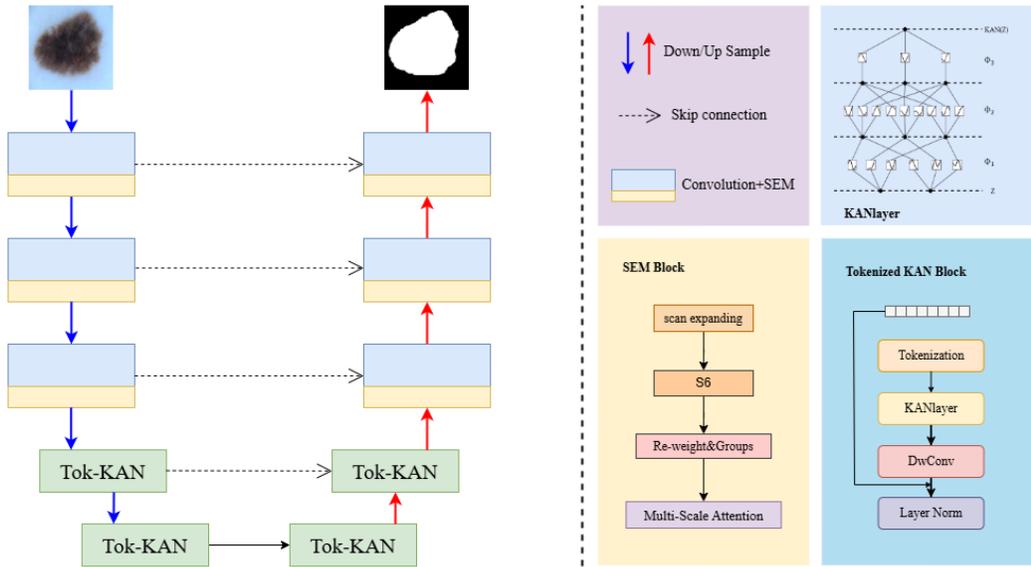

**Figure 1**: Overview of the KM-UNet pipeline. After feature extraction by several convolution blocks and the SEM module in the Convolution Phase, the intermediate maps are tokenized and processed by stacked Tok-KAN blocks in the Tokenized KAN Phase.

## 3. Methodology

**Overview**  Figure 1 illustrates the overall architecture of the proposed KM-UNet. The architecture adopts a three-stage encoder-decoder structure, consisting of the Convolution Phase, the Selective-Scan Efficient Multi-scale (SEM) attention module phase, and the tokenized Kolmogorov-Arnold Network (Tok-KAN Phase). In the encoder, the input image first passes through three convolution operations combined with SEM modules, followed by further processing through two tokenized MLP blocks. The decoder consists of two tokenized KAN blocks, followed by three convolution blocks. Each encoder module reduces the feature resolution by half, while each decoder module doubles the feature resolution. Furthermore, skip connections are integrated between the encoder and decoder to facilitate feature fusion. The number of channels in each module of the Convolution Phase and Tok-KAN Phase is defined by hyperparameters $C_1$ to $C_5$ and $D_1$ to $D_5$, respectively.

**Selective-Scan Efficient Multi-scale (SEM) Attention Module**

The Selective-Scan Efficient Multi-scale (SEM) attention module is designed to enhance the feature extraction and attention mechanism in the encoder and decoder stages of the U-Net network. The module consists of two main parts: the feature extraction part and the attention part.

**Feature extraction**
As shown in Figure 2, the feature extraction part draws inspiration from the SS2D module and improves upon its scanning approach to achieve more efficient feature capture and multi-directional feature extraction. Specifically, the input feature map is unfolded into sequences along multiple directions to capture rich spatial information. The improved scanning method not only includes standard directions such as from top-left to bottom-right and bottom-right to top-left but also introduces an adaptive scanning strategy that flexibly adjusts the scanning order to accommodate different input features.
The expression for each scanning direction is as follows:
$$X^{(\text{dir})} = \text{Scan}(X, \text{direction}) \quad (1)$$
Where direction∈{top-left to bottom-right, top-right to bottom left, bottom-right to top-left, bottom-left to top-right}. The scan result from each direction is passed through the improved S6 block for feature extraction, where the S6 block dynamically adjusts its parameters based on the input to filter and retain the most important features. Subsequently, as illustrated in Figure 2, the Re-weight operation sums and merges the sequences from the four directions, restoring the output image to the same size as the input. The S6 block, derived from Mamba [16], introduces a selective mechanism on top of S4

[17] by adjusting the SSM's parameters based on the input. This enables the model to distinguish and retain pertinent information while filtering out the irrelevant. The pseudo-code for the S6 block is presented in Algorithm 1.

The final feature integration process combines the features from different directions as follows:

$$X' = \sum_{\text{dir}} \text{rMerge}\left(S6\left(X^{(\text{dir})}\right)\right) \tag{2}$$

---

**Algorithm 1** Pseudo-code for S6 block

**Input**: x, the feature with shape [B, L, D] (batch size, token length, dimension)
**Params**: A, the nn.Parameter; D, the nn.Parameter
**Operator**: Linear(.), the linear projection layer
**Output**: y, the feature with shape [B, L, D]
1: $\Delta$, **B**, **C** = Linear(x), Linear(x), Linear(x)
2: $\overline{\mathbf{A}}$ = exp($\Delta \mathbf{A}$)
3: $\overline{\mathbf{B}}$ = $(\Delta \mathbf{A})^{-1}(\exp(\Delta \mathbf{A}) - \mathbf{I}) \cdot \Delta \mathbf{B}$
4: $h_t = \overline{\mathbf{A}} h_t + \overline{\mathbf{B}} x_t$
5: yt = **C**ht + **D**xt
6: y = [y1, y2, · · · , yt, · · · , yL]
7: **return** y

---

**Multi-Scale Attention**

The multi-scale attention module uses two parallel convolutional sub-networks to capture both short-range and long-range spatial dependencies, avoiding the common issue of channel dimensionality reduction. By reshaping the channel dimension of the input feature map into the batch dimension and using different convolution kernels, it ensures efficient computation while preserving key feature information from medical images.

First, the input feature map X with dimensions [B,C,H,W] (where B is the batch size, C is the number of channels, H and W are the height and width) is reshaped to a new form X′, preserving more feature information:

$$X' = \text{reshape}(X) \tag{3}$$

Then, the attention mechanism captures spatial dependencies using two parallel sub-networks. One uses a 1×1 convolution to model local channel interactions, while the other uses a 3×3 convolution to capture broader spatial relationships:

$$X_{\text{cross}} = \text{Conv}_{1\times1}(X) + \text{Conv}_{3\times3}(X) \tag{4}$$

Finally, the outputs of both sub-networks are aggregated to generate the final feature map Y, which combines multi-scale information to improve feature representation:

$$Y = Aggregation(X_{\text{cross}}) \tag{5}$$

This approach captures richer spatial dependencies and contextual information while maintaining low computational complexity, significantly improving feature extraction and model performance for medical image analysis.

### Integration of KAN into U-Net Architecture for Medical Image Segmentations

In this work, we integrate the Kolmogorov-Arnold Network (KAN) as the bottleneck layer in the U-Net architecture to improve feature modeling and enhance the network's interpretability, specifically for medical image segmentation tasks. The proposed U-Net variant, KM-UNet, adopts a three-phase encoder-decoder structure composed of the Convolution Phase, the Selective-Scan Efficient Multi-scale (SEM) attention module phase, and the Tokenized KAN (Tok-KAN) phase. Given an input feature map $Z \in R^{C \times H \times W}$ from the encoder, the KAN layer processes this feature as:

$$Z' = \text{LN}\left(Z + \text{DwConv}(\Phi(Z))\right) \tag{6}$$

where LN denotes layer normalization and DwConv represents depth-wise convolution. This formulation allows the network to effectively capture complex, non-linear dependencies across multi-dimensional features, which is particularly valuable in medical image segmentation, where intricate anatomical structures need to be accurately segmented. By integrating KAN, KM-UNet enhances the model's ability to maintain long-range contextual information and multi-scale feature representation, critical for handling the variability and complexity found in medical images.

KANs, based on the Kolmogorov-Arnold representation theorem, replace the traditional linear transformation matrices used in Multi-Layer Perceptrons (MLPs) with learnable, parameterized activation functions. This not only reduces the parameter overhead but also improves the interpretability of the network, which is vital in medical imaging applications where understanding the model's decision-making process is crucial. The mathematical structure of KANs allows for a more efficient and transparent mapping of features:

$$\text{MLP}(Z) = (W_{K-1} \circ \sigma \circ W_{K-2} \circ \ldots \circ W_1 \circ \sigma \circ W_0)Z \tag{7}$$

In contrast, the KAN layer's structure can be represented as:

$$\text{KAN}(Z) = (\Phi_{K-1} \circ \Phi_{K-2} \circ \ldots \circ \Phi_1 \circ \Phi_0)Z \tag{8}$$

where each $\Phi_i$ consists of learnable activation functions. The output from layer k to layer k+1 is represented as:

$$Z_{k+1} = \Phi_k Z_k \tag{9}$$

This innovative approach, where KANs replace linear transformation matrices, allows the model to achieve superior performance with fewer parameters, making it especially suitable for resource-constrained medical image segmentation tasks. Furthermore, the interpretability of the model is significantly improved, making it a valuable tool for clinicians who need to understand how the model is making its predictions in complex, high-stakes medical scenarios.

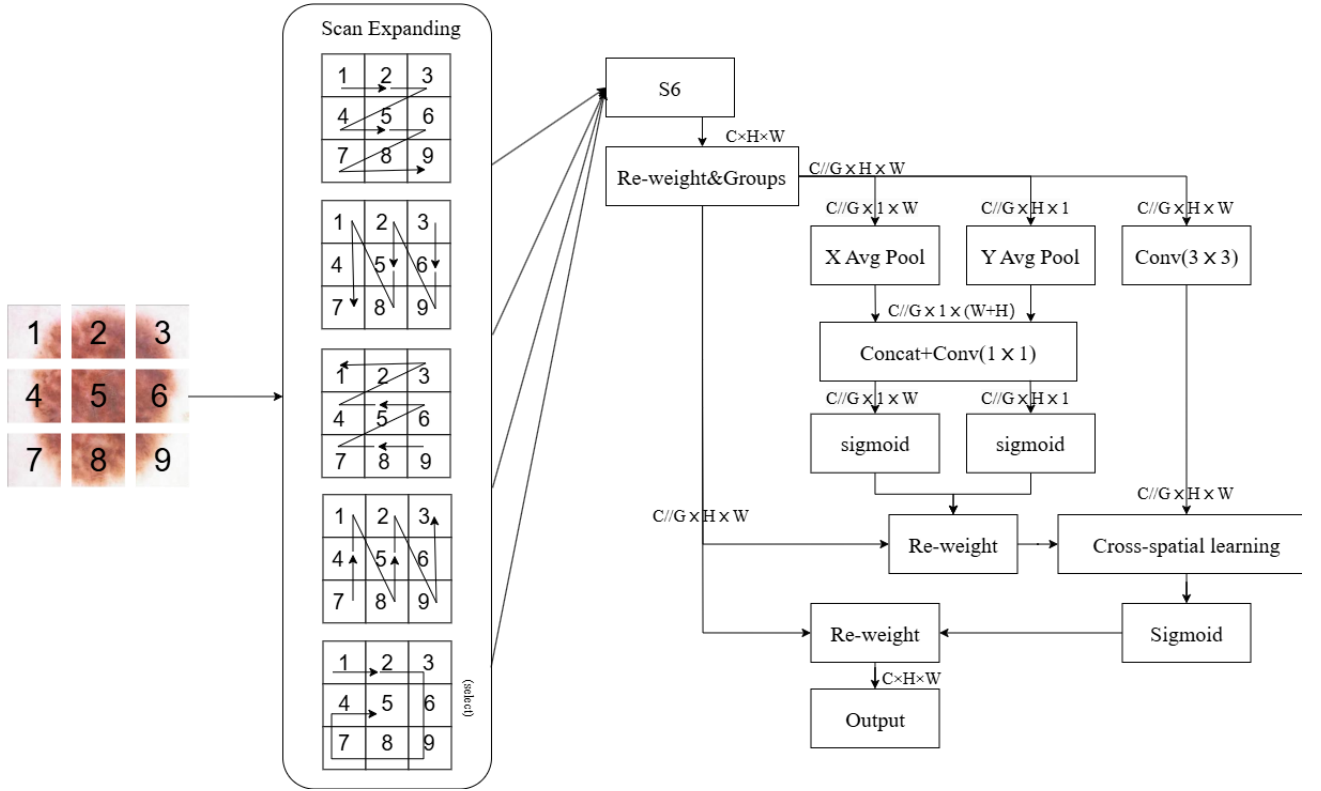

**Figure 2**: Schematic of the SEM Module. The SEM module improves feature extraction and attention mechanisms in U-Net by capturing multi-directional spatial information and refining feature maps for enhanced segmentation.

## 4. Experiments

**Datasets**

We conducted an extensive evaluation of our proposed method using three distinct and heterogeneous datasets, each characterized by unique properties, varying data sizes, and different image resolutions. These datasets are commonly employed in tasks such as image segmentation and generation, providing a comprehensive testing ground to assess the effectiveness and adaptability of our method.
**BUSI Dataset**: The BUSI dataset [25] consists of ultrasound images that depict normal, benign, and malignant breast cancer cases, along with their corresponding segmentation maps. For this study, we used a subset of 647 ultrasound images representing both benign and malignant breast tumors, all uniformly resized to 256 × 256 pixels. This dataset provides a well-rounded collection of images that facilitate the detection and differentiation of various types of breast tumors, offering valuable insights for both medical professionals and researchers.

**GlaS Dataset**: The GlaS dataset [26]comprises 612 standard-definition (SD) frames derived from 31 sequences, with each frame having a resolution of 384 × 288 pixels. These frames were sourced from 23 patients and were collected at the Hospital Clinic in Barcelona, Spain. The data was recorded using devices such as the Olympus Q160AL and Q165L, paired with an Extra II video processor. Following established protocols[34], we used 165 images from this dataset, resized uniformly to 512 × 512 pixels, for our evaluation.

**CVC-ClinicDB dataset** [27]: often referred to as "CVC," is a publicly accessible resource widely used for polyp detection within colonoscopy videos. It contains 612 images with a resolution of 384 × 288 pixels, extracted from 31 distinct colonoscopy sequences. This dataset provides a diverse representation of polyp instances, making it particularly valuable for the development and evaluation of polyp detection algorithms. For consistency across all datasets used in this study, we resized all images from the CVC-ClinicDB dataset to 256 × 256 pixels.

**ISIC17 and ISIC18 Datasets**: The International Skin Imaging Collaboration 2017 and 2018 challenge datasets (ISIC17 and ISIC18) [28][29] are publicly available skin lesion segmentation datasets, containing 2,150 and 2,694 dermoscopy images with corresponding segmentation masks, respectively. In line with prior work [30], we split the datasets in a 7:3 ratio for training and testing. Specifically, the ISIC17 dataset is divided into a training set of 1,500 images and a test set of 650 images, while the ISIC18 dataset comprises a training set of 1,886 images and a test set of 808 images. For both datasets, we conduct detailed evaluations using metrics such as Mean Intersection over Union (mIoU) and Dice Similarity Coefficient (DSC) to assess the performance of our method.

**ImplementationDetails**

We implement KM-UNet using PyTorch on an NVIDIA RTX 4090 GPU. For the BUSI, GlaS, CVC-ClinicDB, ISIC17, and ISIC18 datasets, we set the batch size to 8 and used an initial learning rate of 1e-4, following a cosine annealing learning rate schedule with a minimum learning rate of 1e-5. The Adam optimizer was utilized for training, and the loss function combined binary cross-entropy (BCE) and Dice loss for robust performance. Each dataset was randomly split into 80% for training and 20% for validation. The model was trained for a total of 300 epochs, and all results were averaged over three independent runs to ensure reliability. Only basic data augmentations, including random rotation and flipping, were applied. We evaluated the segmentation results both qualitatively and quantitatively using metrics such as Mean Intersection over Union (IoU) and Dice Similarity Coefficient (F1 Score).

**Table 1**: Comparison with state-of-the-art segmentation models on three heterogeneous medical scenarios.

| Methods | BUSI | | GlaS | | CVC | | ISIC17 | | ISIC18 | |
|---|---|---|---|---|---|---|---|---|---|---|
| | IoU ↑ | F1 ↑ | IoU ↑ | F1 ↑ | IoU ↑ | F1 ↑ | IoU ↑ | F1 ↑ | IoU ↑ | F1 ↑ |
| U-Net | 57.22 | 71.91 | 86.66 | 92.79 | 83.79 | 91.06 | 76.98 | 86.99 | 77.86 | 87.55 |
| Att-Unet | 55.18 | 70.22 | 86.84 | 92.89 | 84.52 | 91.46 | 76.37 | 86.22 | 78.43 | 87.91 |
| U-Net++ | 57.41 | 72.11 | 87.07 | 92.96 | 84.61 | 91.53 | 78.58 | 86.35 | 78.31 | 87.83 |
| U-NeXt | 59.06 | 73.08 | 84.51 | 91.55 | 74.83 | 85.36 | 82.47 | 89.69 | 82.90 | 90.38 |
| Rolling-UNet | 61.00 | 74.67 | 86.42 | 92.63 | 82.87 | 90.48 | 82.14 | 90.22 | **84.15** | **91.13** |
| U-Mamba | 61.81 | 75.55 | 87.01 | 93.02 | 84.79 | 91.63 | 81.47 | 89.07 | 80.92 | 89.49 |
| **KM-UNet（Ours)** | **65.42** | **78.79** | **87.51** | **93.27** | **85.01** | **91.79** | **84.05** | **91.15** | 83.84 | 91.00 |

Table 2: Overall comparison with state-of-the-art segmentation models w.r.t. efficiency and segmentation metrics.

| Methods | Average Seg. | | Efficiency | |
|---|---|---|---|---|
| | IoU ↑ | F1 ↑ | Gflops | Params (M) |
| U-Net | 76.50 | 86.06 | 524.2 | 34.53 |
| Att-Unet | 76.27 | 85.74 | 533.1 | 34.9 |
| U-Net++ | 77.20 | 86.16 | 1109 | 36.6 |
| U-NeXt | 76.75 | 86.01 | 4.58 | 1.47 |
| Rolling-UNet | <u>79.32</u> | <u>87.83</u> | 16.82 | 1.78 |
| U-Mamba | 79.20 | 87.75 | 2087 | 86.3 |
| **KM-UNet（Ours)** | **81.17** | **89.20** | 17.66 | 7.35 |

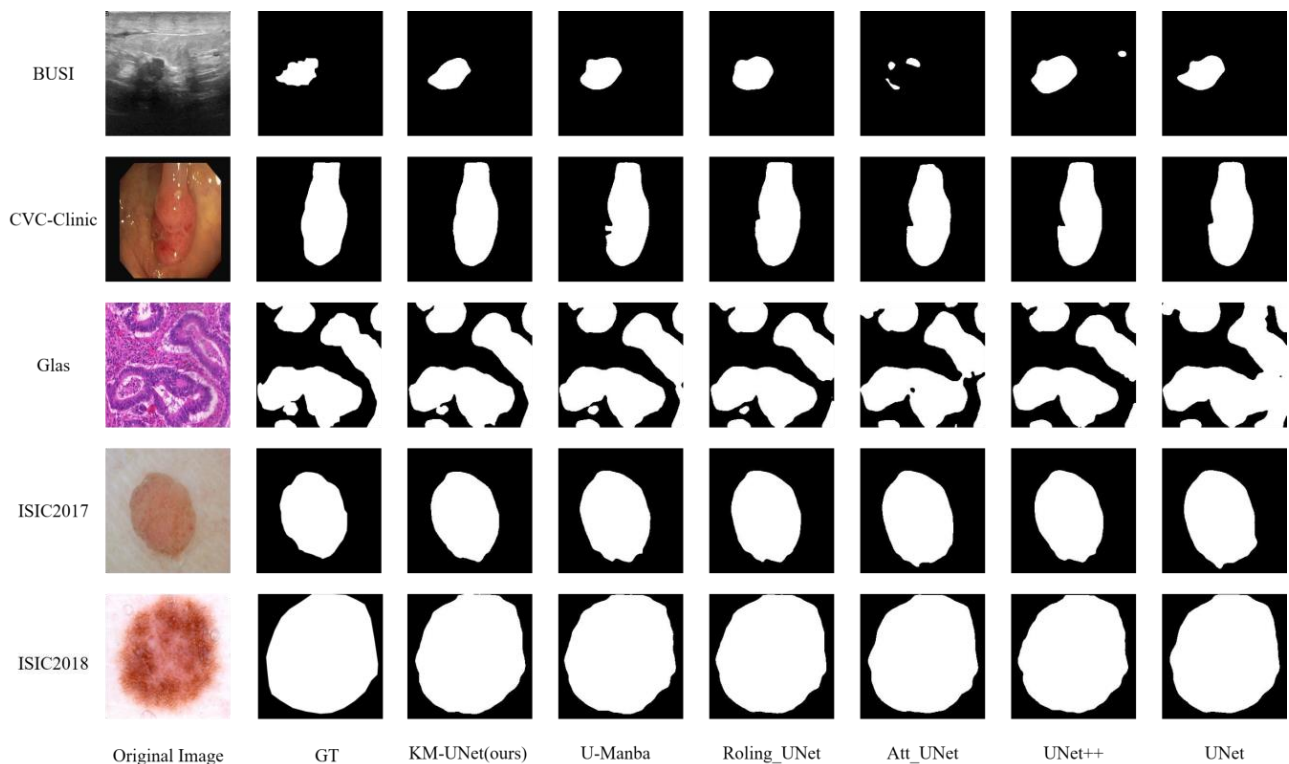

Figure 3: Visualized segmentation results of the proposed KM-UNet against other state-of-the-arts over three heterogeneous medical scenarios.

## Main results

In the comparative experiments, KM-UNet was evaluated against several existing segmentation methods using standard metrics, Intersection over Union (IoU) and F1 score, to assess accuracy and robustness across diverse datasets. Table 1 presents the segmentation performance, while Table 2 highlights the trade-off between accuracy and efficiency.
We compared KM-UNet with traditional CNN-based models such as U-Net [31] and UNet++ [32], attention-based methods like Att-UNet [33], efficient transformer variants like U-Mamba [17], and advanced MLP-based models including U-NeXt [20] and Rolling-UNet. The results show that KM-UNet achieves the highest IoU and F1 scores on most datasets and strikes an excellent balance between segmentation accuracy and efficiency, as shown in Table 2. Notably, it achieves an average IoU of 80.45% and F1 score of 88.63%, with a relatively small parameter count (7.35M) and computational cost (17.66 Gflops).

The integration of SEM model in KM-UNet is instrumental to its performance. SEM effectively models global context and long-range dependencies, addressing the limitations of CNN-based models in capturing global information. By leveraging S6 model, KM-UNet preserves fine boundary details, reduces false predictions, and maintains computational efficiency, making it well-suited for challenging datasets such as ISIC17 and BUSI.

Compared to models like U-Net and UNet++, KM-UNet demonstrates superior context modeling and detail preservation. Against attention-based and MLP-based models, including U-Mamba, KM-UNet offers a better balance between segmentation accuracy and efficiency. Its performance across both accuracy and computational cost, as shown in Tables 1 and 2, highlights its overall effectiveness in medical image segmentation.

**Explainability**

This study explores the role of the KAN layer in improving model interpretability. By comparing attention heatmaps, as shown in Figure 7, when the KAN layer is not used (first column), the model fails to accurately locate key regions, resulting in a low IoU, indicating insufficient overlap between the activation areas and the ground truth masks. In contrast, after integrating the KAN layer (second column), the model can precisely locate the target boundaries, and the generated activation areas align better with the ground truth masks, leading to a significant improvement in the IoU. This result demonstrates that the KAN layer, by introducing an attention mechanism, enhances the model's focus on key regions, improving both prediction accuracy and interpretability. This improvement is consistent with the performance observed in other studies on the KAN layer, validating its effectiveness in increasing model transparency.

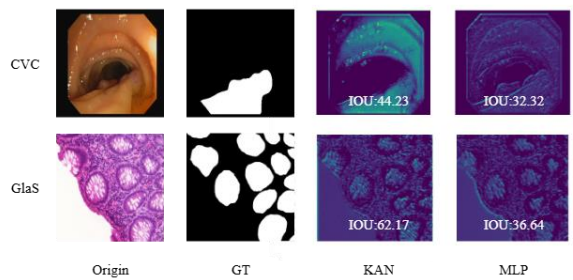

**Figure 7**: Explainability of KAN with channel activation.

## 5. Conclusions and Future Works

The proposed KM-UNet framework demonstrates significant performance improvements in medical image segmentation tasks. By introducing the Selective-Scan Efficient Multi-scale (SEM) module, Kolmogorov-Arnold Network (KAN), and advanced learning rate scheduling strategies, substantial enhancements over the traditional U-Net model have been achieved. Ablation studies show that the SEM module, which combines feature extraction and attention mechanisms, effectively improves segmentation accuracy, with a 2%-3% increase in IoU and F1-score. Additionally, a comparison of different learning rate scheduling strategies reveals that the cosine annealing learning rate scheduler outperforms the fixed learning rate configuration, resulting in slightly better experimental outcomes. The substitution of the traditional MLP network with the KAN network significantly boosts global feature modeling capability, leading to improved segmentation results.

However, there are still areas that need further optimization. Although the SEM module shows excellent performance, its computational complexity remains relatively high, and future work could focus on designing more efficient feature extraction and attention mechanisms to reduce computational overhead. Moreover, while KM-UNet is currently applied to medical image segmentation, it could be extended to other domains such as remote sensing or video segmentation, which may require adjustments to the module structure to suit the specific demands of these tasks. As deep learning models continue to grow in complexity, reducing the model's computational cost and parameter count while maintaining accuracy becomes a major challenge. Lightweight techniques such as quantization and pruning will be key areas for future optimization. Furthermore, incorporating multimodal medical imaging data, such as CT and MRI, into KM-UNet could enhance the model's ability to identify complex pathologies and improve segmentation accuracy. Lastly, as datasets continue to expand, federated learning and transfer learning could further enhance KM-UNet's performance on smaller datasets. Exploring multi-task learning and transfer learning techniques will be crucial for improving KM-UNet's generalization across various scenarios.